An accurate and efficient method for calculating fluid exchange between fractures and matrix with a non-conforming mesh

Mark McClure

McClure Geomechanics LLC

**Abstract**

I propose the 1D Subgrid Method for calculating fluid exchange between fractures and matrix with a non-conforming mesh. The method is demonstrated on linear and radial flow problems. The method has negligible computational cost and is accurate at all levels of mesh refinement.

**1. Introduction**

Calculation of fracture-matrix fluid exchange is a challenging aspect of discrete fracture network simulation. Flow can be calculated with a conforming or nonconforming mesh. With a conforming mesh, standard numerical techniques can be used to calculate fluid exchange (Karimi-Fard et al., 2004). However, with complex fracture geometries, construction of a conforming mesh can be very challenging, especially in 3D. If fracture propagation occurs, then conforming meshes require remeshing. Techniques using a nonconforming mesh are useful because they avoid the complexities of unstructured discretization and do not require remeshing (Li and Lee, 2008; Norbeck et al., 2015; Karvounis and Jenny, 2016).

Both structured and unstructured methods using the finite volume method require a sufficiently refined mesh to be accurate. Leakoff from a fracture in very low permeability rock can only be calculated accurately if the mesh is highly refined near the fracture. This increases computational cost and requires remeshing in problems with propagating fractures.

I propose the 1D Subgrid Method to resolve these difficulties. The 1D Subgrid Method allows accurate calculation of leakoff rate with coarse nonconforming meshes, even if permeability is very low. I demonstrate the method's accuracy with two test cases – a linear flow and radial flow example. The method can describe spherical flow or any other leakoff geometry. The computational cost of the 1D Subgrid Method is negligible, compared with a standard finite volume solution using the same mesh.

**2. Methods**

Fluid flow between the fracture and matrix element is calculated using the formula:

$$q_p = \frac{\rho_p k_{rp}}{\mu_p} T(P_f - P_m^*), \tag{1}$$

where $q_p$ is mass flow rate of phase $p$, $\rho_p$ is density of phase $p$, $k_{rp}$ is the relative permeability of phase $p$, $\mu_p$ is viscosity of phase $p$, $T$ is a transmissibility factor, $P_f$ is the fluid pressure in the fracture, and $P_m^*$ is an effective matrix fluid pressure. The $\frac{\rho_p k_{rp}}{\mu_p}$ terms are evaluated with appropriate choice of upwinding.

If $P_m^*$ is set equal to the fluid pressure in the matrix element, Equation 1 will not yield accurate leakoff calculations if the mesh is coarse (relative to the pressure diffusion distance over a reference timescale), regardless of the value chosen for $T$. The 1D Subgrid Method is based on making careful choice of $T$ and $P_m^*$ to alleviate this inaccuracy.

For each fracture-matrix connection, a one-dimensional subgrid is used to calculate $P_m^*$ and $T$. A fracture element may not be solely confined within a single matrix element, and so multiple subgrids may be constructed for each fracture element. The procedure for generating the subgrid is:

1. Generate a logarithmically spaced one-dimensional grid. The length of the subgrid element adjacent to the fracture should be no more than $\sqrt{4Dt_m}$, where $D$ is the hydraulic diffusivity in the matrix and $t_m$ is a small characteristic timescale. The gridblock length is increased geometrically away from the fracture, until reaching a distance that is greater than the largest dimension of the matrix element. This grid is used for calculating $P_m^*$.
2. Generate a coarse-scale, uniformly spaced, one-dimensional grid with the same length as the logarithmically spaced subgrid. This grid is used temporarily for calculating the cross-sectional area of the subgrid elements in Step #1.
3. Split the matrix element into a series of smaller subblocks. For each subblock, identify which fracture element within the matrix element is closest to that subblock. Calculate the distance from the subblock to the fracture element. Assign the volume of the subblock to the coarse-scale 1D grid element that corresponds to that distance. Each subblock should be assigned only to the coarse-scale subgrid of the fracture element that is closest to it.
4. For each coarse-scale subgrid element, calculate the total volume of associated matrix subblocks. Divide this volume by the element length to calculate the element's cross-sectional area.
5. Truncate the coarse-scale subgrid so that the total volume of the remaining elements is 50-75% of the total volume of the matrix subblocks assigned to the fracture element.
6. Interpolate the coarse-scale subgrid cross-sectional areas onto the logarithmically spaced subgrid from step #1. In the interpolation, the cross-sectional area on the side adjacent to the fracture is set equal to two times the area of the portion of the fracture element surface area that is located within the matrix element. The factor of two accounts for leakoff from both sides of the element.

To calculate $P_m^*$, the diffusivity equation is solved numerically within each 1D subgrid. The diffusivity equation is:

$$\frac{dP}{dt} = \frac{k}{\phi c_t \mu} \frac{d^2 P}{dx^2}. \tag{2}$$

The 1D grid can be discretized using finite difference into a linear tridiagonal system, which can be solved very efficiently to calculate $P$ in each subblock. For boundary conditions, the pressure on the fracture side of the subgrid is set to $P_f$ and the pressure on the matrix side of the subgrid is set to $P_m$. The properties in $\frac{k}{\phi c_t \mu}$ are selected from the matrix block.

The value of $P_m^*$ in Equation 1 is equal to the pressure of the element in the 1D subgrid directly adjacent to the fracture element. The value of $T$ in Equation 1 is equal to the area divided by the half-length of that subgrid element.

If a new fracture element is created in the matrix block, then the 1D subgrids for all fracture-matrix connections of the block must be recalculated. The pressure distributions from the original subgrids are interpolated onto the pressure distribution of the new subgrids.

Multiphase leakoff calculations can be accomplished with the same procedure. In this case, Equation 2 is replaced with multiphase flow equations. The finite difference calculation is more complex because it requires solution to a nonlinear system of equations and tracking of multiple components.

Even if the overall simulation is multiphase, it is possible to use the single-phase version of Equation 2 for calculating $P_m^*$ in Equation 1. This is an approximation that does not capture all the details of multiphase leakoff.

The most convenient way to include gravity is to add a hydraulic head adjustment to the matrix pressure boundary condition in the 1D solution to the diffusivity equation.

## 3. Results and Discussion

The algorithm was implemented in a fracturing and reservoir simulator. The results from two test problems are presented below. In the first, constant rate injection is performed from an infinite conductivity fracture that extends the full length of the matrix mesh so that the flow geometry is one-dimensional. In the second, constant rate injection is performed from a short infinite conductivity fracture so that flow geometry is radial. The simulations are single phase and use a permeability of 0.001 md.

Figure 1 shows the linear flow calculation using standard finite difference with a fine mesh. The fine mesh uses logarithmic spacing, with greater refinement near the fracture. The log-log derivative plot indicates a 1/2 slope, as expected (Horne, 1995). Figure 2 shows the result for the 1D flow case with a coarse mesh and a standard finite difference algorithm. The solution is extremely inaccurate and only approaches a 1/2 slope towards the end of the simulation. The solution is inaccurate as long as $\sqrt{4Dt}$ is small relative to the matrix block size. Figure 3 shows the result using the same coarse mesh, but using the 1D Subgrid Method. The result is very close to the fine-mesh solution. Figure 4 shows a result with the 1D Subgrid Method using an extremely coarse mesh - a single matrix element for the entire problem domain. The solution is accurate, even in this extreme case.

Figure 5 shows the radial flow calculation with standard finite difference and a fine mesh. The log-log derivative plot indicates a 0 slope, as expected. Figure 6 shows the radial flow result with a coarse mesh and the 1D Subgrid Method. Figure 7 shows the radial flow result with the 1D Subgrid Method and the entire problem domain meshed with a single element. In both cases, the result with the 1D Subgrid Method is accurate. These results confirm that the 1D Subgrid Method accurately describes radial flow, not just linear flow. In fact, the method is general and capable of describing any flow geometry, such as spherical, hemispherical, or hemiradial. Figure 8 shows the result with the 1D Subgrid Method and the same fine mesh used in Figure 5. The results in Figures 5 and 8 are identical, verifying that the 1D

Subgrid Method converges to the exact solution with mesh refinement. The simulations confirm that the 1D Subgrid Method is robust for all levels for mesh refinement, from very coarse to very fine.

The derivative curve in the results with the 1D Subgrid Method has a few wiggles caused by inaccuracy. The wiggles occur approximately when the pressure diffusion length reaches the end of the 1D subgrid. This indicates difficulty in transitioning from flow behavior dominated by the 1D subgrid to flow behavior dominated by the overall problem mesh. Probably, the treatment of the matrix-side boundary condition in the 1D calculation could be improved to reduce or eliminate these wiggles. Because of these artifacts, the 1D Subgrid Method probably should not be used for simulations intended specifically for analysis of pressure transient behavior, in which the behavior of the pressure derivative will be carefully scrutinized. However, for typical applications such as hydraulic fracture modeling, the slight imperfections evident in the pressure derivative will have a negligible effect on the overall results of the simulation. It should be noted that taking the derivative with respect to the logarithm of time tends to exaggerate changes in the transient behavior.

Unless the 1D Subgrid Method is implemented with multiphase flow, it will not capture detailed multiphase processes, such as the accumulation of injection fluid near the fracture, which could reduce relative permeability during subsequent production of fluid from the formation.

The 1D Subgrid Method calculates exchange between matrix and fracture elements independently of other matrix-fracture pairs. This could potentially cause inaccuracy in complex cases, particularly those involving multiple fracture in the same matrix element.

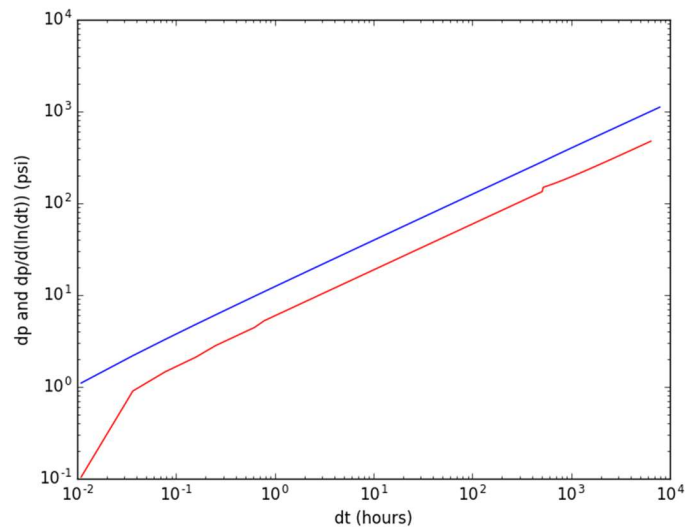

Figure 1: One-dimensional flow calculation using standard finite difference and a fine mesh.

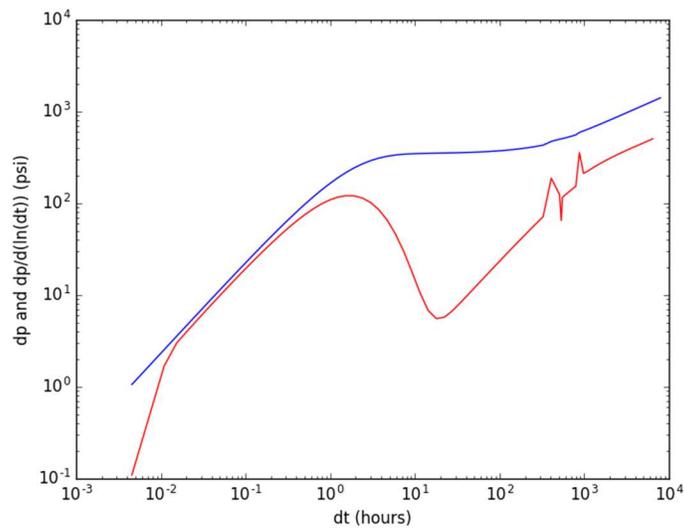

Figure 2: One-dimensional flow calculation using standard finite difference and a coarse mesh.

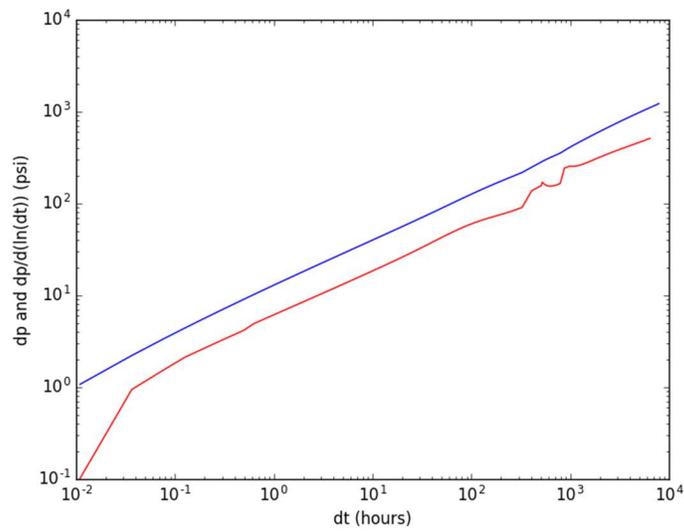

Figure 3: One-dimensional flow calculation using the 1D Subgrid Method and a coarse mesh.

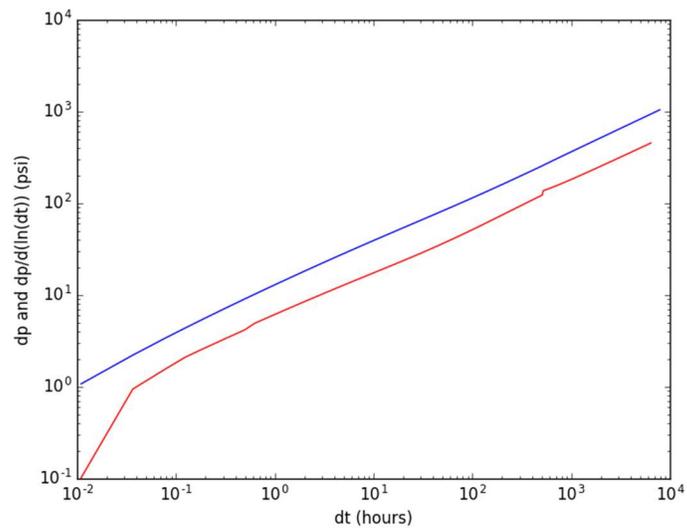

Figure 4: One-dimensional flow calculation using the 1D Subgrid Method and a very coarse mesh.

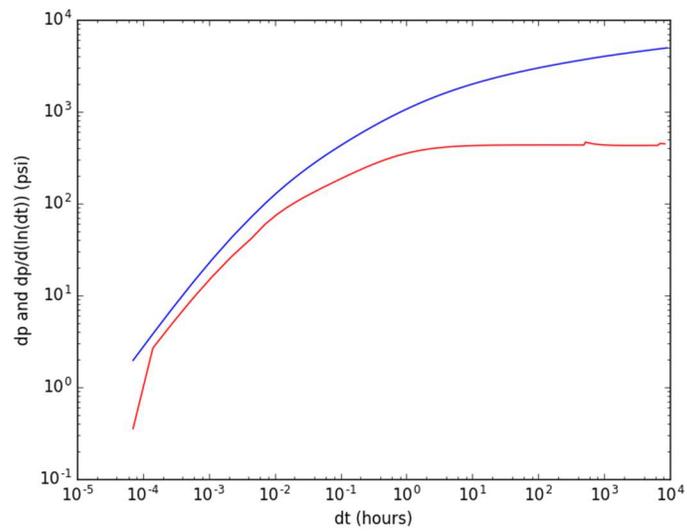

Figure 5: Radial flow calculation using standard finite difference and a fine mesh.

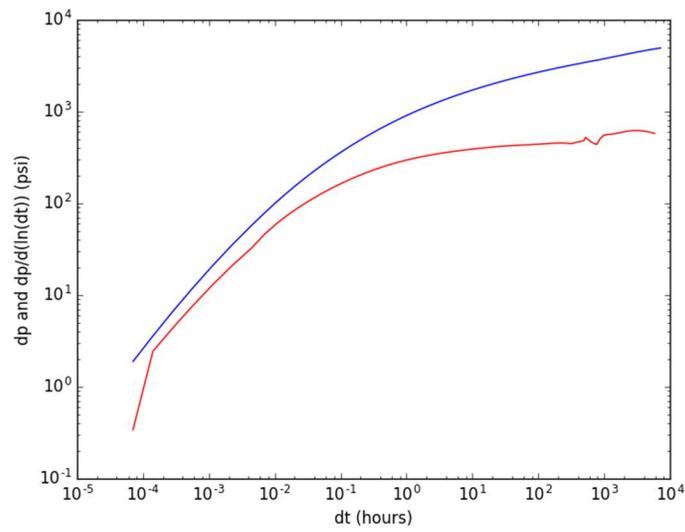

Figure 6: Radial flow calculation using the 1D Subgrid Method and a coarse mesh.

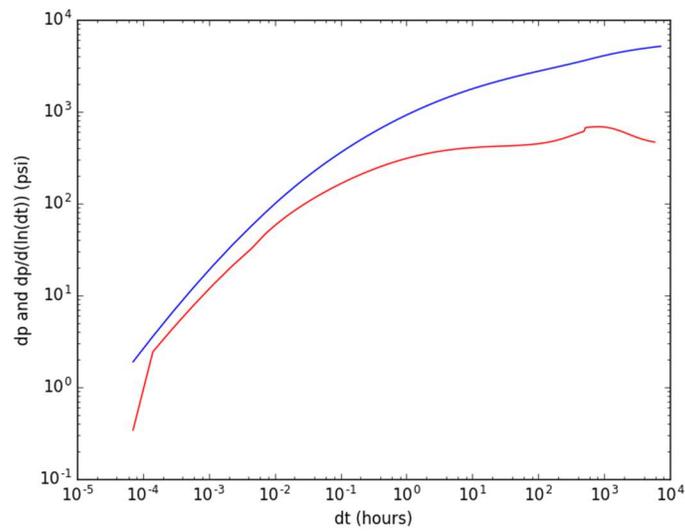

Figure 7: Radial flow calculation using the 1D Subgrid Method and a very coarse mesh.

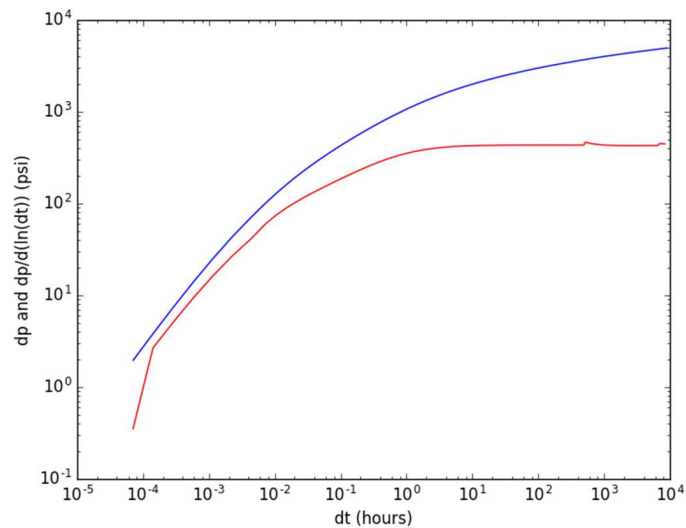

Figure 8: Radial flow calculation using the 1D Subgrid Method and a fine mesh.

**Conclusions**

The 1D Subgrid Method is able to accurately calculate fluid exchange between fracture and matrix elements in a coarse nonconforming mesh. Accuracy is retained at all levels of mesh refinement, and it can handle any leakoff geometry. The computational cost is negligible. Simulations using the method have minor imperfections in the pressure derivative curve.

**Acknowledgements**

Thank you to Jack Norbeck and Charles Kang for informally reviewing this manuscript.